\documentstyle[prl,aps,epsf,12pt]{revtex}
\begin{document}
\draft
 
\title{Resonant neutron scattering on the high Tc cuprates\\ and
$\pi$ and $\eta$ excitation of the $t$-$J$ and Hubbard models}
\author{Eugene Demler$^{1}$ and Shou-Cheng Zhang$^{2}$}
\address{
$^{1}$Institute for Theoretical Physics, University of California, 
Santa Barbara, CA 93106
}
\address{$^{2}$
Department of Physics,
Stanford University
Stanford, CA 94305
}
\date{\today}
\maketitle
\begin{abstract}
We review the explanation for resonant neutron scattering experiments in
$YBa_2Cu_3O_{6+\delta}$ and $Bi_2 Sr_2 Ca_2 O_{8+\delta}$ 
materials from the point of view of triplet excitation in 
the particle-particle channel, the $\pi$ excitation. Relation
of these resonances to the superconducting condensation energy 
and their role in stabilizing the superconducting state is discussed.
Due to the superconducting fluctuations, the $\pi$ resonance may
appear as a broad peak above Tc. 
Analogue problem to the $\pi$ excitation for the case of 
$s$-wave pairing, the $\eta$ excitation, is considered in the strong coupling 
limit with an emphasis on the resonance precursors in a
state with no long range order. 
\end{abstract}

Inelastic neutron scattering (INS) on optimally doped $YBa_2Cu_3O_{7}$ revealed a
striking resonance at the commensurate wavevector $Q=(\pi/a,\pi/a)$ and
energy $41~meV$ \cite{optimal}. This resonance appears in the spin-flip
channel only, therefore it is of magnetic origin and not due to
scattering by phonons. The most remarkable feature of this resonance
is that it appears only below Tc and therefore tells us about enhanced
antiferromagnetic fluctuations in the superconducting state of this
material. Similar resonances have later been found for the underdoped
$YBa_2Cu_3O_{6+\delta}$ at smaller energies but the same wavevector of commensurate
antiferromagnetic fluctuations $(\pi/a,\pi/a)$\cite{underdoped}.  A new
feature of the resonances in underdoped materials is that they no
longer disappear above Tc but have precursors appearing at some
temperature above the superconducting transition temperature.  
Recently, resonant peaks below Tc have also been observed in INS
experiments on $Bi_2 Sr_2 Ca_2 O_{8+\delta}$ materials \cite{BSCO}.

Several theories have been suggested to explain the observed
resonances\cite{theories}. One of the proposals\cite{prl95} was to identify
them with a $\pi$-excitation, a triplet excitation at momentum $(\pi/a,\pi/a)$
in the particle-particle channel :~ 
$
\pi^{\dagger} = \sum_p ( cos p_x - cos p_y ) 
	c_{p+Q \uparrow}^{\dagger} c_{-p\uparrow}^{\dagger}
$. 
This excitation is a well-defined collective mode for a wide range of
lattice Hamiltonians, including the Hubbard Hamiltonian\cite{meixner} and
the $t$-$J$ model\cite{prl95,prb,eder}.
A simple way to visualize the $\pi$-excitation is to think of a
spin-triplet pair of electrons sitting on the nearest sites, having
the center of mass momentum $(\pi/a,\pi/a)$ and the same relative
wavefunction as a $d$-wave Cooper pair. When the system acquires a
long-range superconducting order the $d$-wave Cooper pairs may be
resonantly scattered into the $\pi$ pairs and this gives rise to the
sharp resonance seen in neutron scattering. 
Motivated by a recent proposal of Scalapino and White\cite{scalapino},
we showed that last argument may be
enhanced and we argued that it may be energetically favorable for
the system to become superconducting, so that the $\pi$ channel may
contribute to the spin fluctuations and the system can lower its
antiferromagnetic exchange energy\cite{nature}.  In this way, the $\pi$
excitation may be ``promoted'' from being a consequence of the
$d$-wave superconductivity to being its primary cause. This argument
is similar to the argument for the stabilization of the A phase of
superfluid helium-3, where spin fluctuations are enhanced in the A
phase relative to the B-phase\cite{helium}.  
So the effect of the spin fluctuation
feedback may lead to stabilization of the A phase relative to the B
base, which would always be favored in weak coupling BCS analysis. The
hypothesis of the $\pi$ excitation being the major driving force of
the superconducting transition is supported by comparison of the
superconducting condensation energy and the change in the
antiferromagnetic exchange energy due to the appearance of the
resonance. For optimally doped $YBa_2Cu_3O_7$ the resonance lowers the
exchange energy by $18~K$ per unit cell \cite{nature}, whereas the
condensation energy for this material is $12~K$ per unit
cell\cite{loram}. So the resonance by itself is sufficient to account
for the superconducting condensation energy. At finite temperature
there also seems to be a connection between the resonance intensity
and the electronic specific heat, giving another indication of the important role 
played by the resonance in the thermodynamics of superconducting 
transition (see H. Mook's article in these
proceedings). Another important aspect of the $\pi$ excitation is
that it allows to unify the spin $SU(2)$ and charge $U(1)$ symmetries
into a larger $SO(5)$ symmetry and suggests a new effective low energy
Lagrangian for the description of the high $T_c$ cuprates
\cite{science}. From the point of view of $SO(5)$ symmetry $\pi$
operator is one of the generators of the symmetry and resonances
observed in neutron scattering are pseudo-Goldstone bosons of broken
symmetry.

\begin{figure*}[h]
\centerline{\epsfysize=4.0cm 
\epsfbox{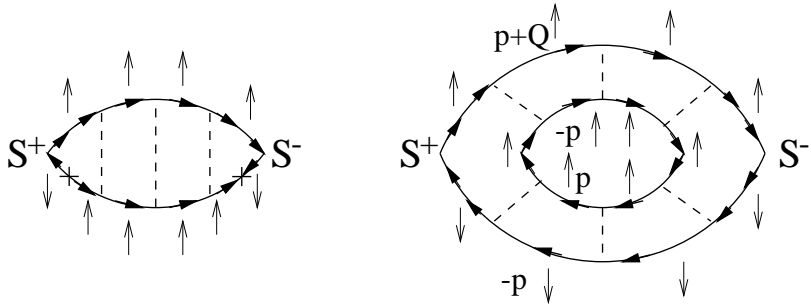}
}
\caption{Feynmann diagram for the $\pi$ resonance below $T_c$ contrasted with
the diagram above $T_c$.}
\label{fig1}
\end{figure*}

Discussion of the $\pi$ excitation in the state with long range
superconducting order has been given within gauge invariant RPA
formalism in \cite{prb} and from the point of view of $SO(5)$
symmetry in \cite{science}. However thorough understanding of
precursors of this excitation in the normal state of underdoped
materials is still lacking. In reference \cite{scz_proc} it was
pointed out that the most likely origin of these precursors is the
existence of strong superconducting fluctuations in the pseudogap
regime of underdoped cuprates \cite{phase_fluct}, 
an assumption phenomenologically supported
by correlations between the temperature at which the resonance
precursors appear and broadening of the singularity in the electronic
specific heat above $T_c$. In weak
coupling precursor of the $\pi$ resonance have been identified with a
process in which a $\pi$-pair and a preformed Cooper pair propagate in
opposite directions \cite{scz_proc} (see Figure \ref{fig1}), 
however quantitative analysis of such process is very difficult. 

In this article we would like to point 
out that there is an analogue problem to
the $\pi$-excitation, $\eta$ excitation of the negative $U$ Hubbard
model\cite{eta,ijmp}, which allows a detailed study in the strong coupling limit,
including the analysis of resonance precursors in a state without long
range order.
So in the remaining part of this article we will review
the $\eta$ excitation, pseudospin $SU(2)$ symmetry that it gives rise
to \cite{zhang-yang} and discuss the strong coupling limit of the negative $U$
Hubbard model.

As originally suggested by Yang and Zhang \cite{zhang-yang} the negative $U$ Hubbard
model (we consider $d=3$ case) 
\begin{eqnarray}
{\cal H} = -t \sum_{\langle ij \rangle \sigma} c_{i\sigma}^{\dagger} c_{j\sigma}
+ U \sum_i ( n_{i \uparrow} - \frac{1}{2} ) ( n_{i \downarrow} - \frac{1}{2} ) 
- \mu  \sum_{ i \sigma}  c_{i\sigma}^{\dagger} c_{i\sigma}
\end{eqnarray}
at half filling has $SO(4)$ symmetry where in addition to the usual
spin $SU(2)$ symmetry the system possesses pseudospin $SU(2)$
symmetry, generated by operators
$\eta^{\dagger} = \sum_p c_{p+Q\uparrow}^{\dagger} c_{-p\downarrow}^{\dagger}$,
~$ \eta_0 = \frac{1}{2} ( N_e - N )$, and
$\eta = \left( \eta^{\dagger} \right)^{\dagger}$,
where $N_e$ is the total number of electrons and $N$ is the total number of sites.
The $\eta$ operator that was introduced is similar to the $\pi$ operator that
we discussed earlier in that it acts in the particle-particle channel and has momentum 
$Q=(\pi/a,\pi/a)$, however it is spin singlet operator and creates pairs of electrons
where electrons sitting on the same site. While the $SO(5)$ symmetry of the $t$-$J$ model
is related to the competition between the antiferromagnetic and d-wave superconducting orders
in this system, the pseudospin $SU(2)$ symmetry relates CDW
and s-wave superconductivity in the negative $U$ Hubbard model. At half filling when 
$\mu=0$ the symmetry is exact and there is a degeneracy between the CDW state and state 
with superconducting order.
Away from half-filling the symmetry is explicitly broken and superconducting state 
is energetically 
more favorable than the CDW state. The
$\eta$ operator no longer commutes with the Hamiltonian but satisfies
$
\left[ {\cal H}, \eta^{\dagger} \right] = - 2 \mu \eta^{\dagger}
$.
Therefore, when it acts on the ground state it creates an 
exact excited state at the energy of $ - 2 \mu$. In the normal state such $\eta$ 
excitation does not contribute to the density fluctuation spectrum, since it is charge 2 operator,
however when the system becomes superconducting and there is a finite probability
of converting particles into holes the $\eta$ resonance shows as a resonance in the
density fluctuation spectrum (one can also think of this as resonant scattering of 
s-wave Cooper pairs into $\eta$ pairs). 
Another way to see how the $\eta$ channel gets coupled
to the density channel below $T_c$ is to notice that there is an exact commutation relation
$
\left[ \eta^{\dagger}, \rho_Q \right] =  \Delta 
$
where $ \rho_Q = \sum_{p\sigma} c_{p+Q\sigma} c_{p\sigma} $ is density operator at momentum $Q$
and $\Delta$ is a superconducting s-wave order parameter. In the superconducting state the right
hand side of this relation acquires an expectation value, so $\rho_Q$ and $\Delta$ become conjugate variables
and resonant peak in one of them shows up as a resonance in the other\cite{eta}.

The question that we want to address is what happens if we do not 
have long range order but
only strong superconducting fluctuations. Can they lead to 
precursors of the $\eta$ resonance in the
density-density correlation 
function and what is the form
of such precursors. The limit that we want to consider is strong 
coupling limit when $U>>t$. 
We perform a particle-hole transformation on a bipartite lattice
\begin{eqnarray}
c_{i\uparrow} \rightarrow c_{i\uparrow} \hspace{1cm}
c_{i\downarrow} \rightarrow \left\{ \begin{array}{l} c_{i\downarrow}^{\dagger}, i \in A, \\
			-	c_{i\downarrow}^{\dagger}, i \in B \end{array} \right\}
\end{eqnarray}
where $A$ and $B$ denote two sublattices.
This transformation maps the negative $U$ Hubbard model at finite doping ( i.e. finite $\mu$ ) into positive
$U$ Hubbard model at half-filling but in the presence of magnetic field along 
$z$-axis. Pseudospin $SU(2)$ symmetry goes into the spin $SU(2)$ symmetry and 
$\eta$ resonance becomes a Larmor resonance in the transverse spin channel.
Particle-hole transformation
maps superconducting order parameter into antiferromagnetic order
parameter in the $x$-$y$ plane and a CDW order parameter into antiferromagnetic order
parameter in $z$ direction. The advantage of doing such particle-hole transformation
is that strong coupling limit of the positive
$U$ Hubbard model at half filling is well known. It is the Heisenberg
model with the nearest neighbor exchange interaction $J=4t^2/U$
\cite{assa}. So the effective Hamiltonian in strong 
coupling is 
\begin{eqnarray}
{\cal H} = J \sum_{\langle ij \rangle} {\bf S}_i {\bf S}_j + H_z \sum_i S_i^z 
\label{heisenberg}
\end{eqnarray}
with $H_z = 2 \mu$.
At $T=0$ Heisenberg model in external field will develop a long
range antiferromagnetic order in the plane perpendicular to the direction
of the applied field ( $xy$ plane in our case, which corresponds 
to the superconducting order in the original negative $U$ model ).
When this happens, $N_{\pm}$ get expectation values and
 we can use the commutation relation
$
\left[ S_{\pm}, N_z \right] = \pm N_{\pm}
$
to show that Larmor resonance, which was originally present in $S_{\pm}$ channels only,
will appear in the $N_z$-$N_z$ correlation function as well. However we are interested
in the regime when we have  only finite range antiferromagnetic correlations  
(although strong)
and we want to find possible precursors of the Larmor resonance in the $N_z$ channel. 

\begin{figure*}[h]
\centerline{\epsfysize=4.0cm 
\epsfbox{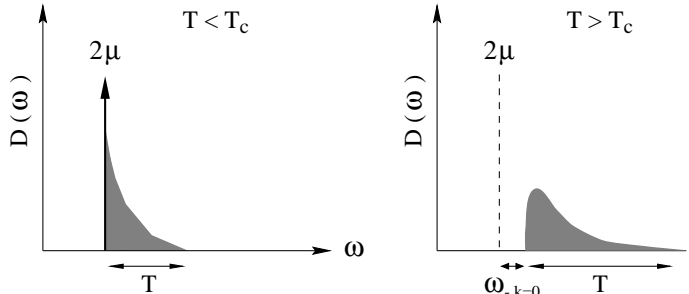}
}
\caption{ 
Density response function at momentum $Q=(\pi/a,\pi/a)$ 
for the negative-$U$ Hubbard model in the large $U$ limit.
For $T<T_c$~ $D(\omega)$ has a $\delta$-peak at $\omega=2 \mu$ ($\eta$ resonance)
and a broad feature starting
at the same energy. For $T>T_c$ there is only a broad peak at the energy slightly
higher then $2\mu$.}
\label{fig2}
\end{figure*}

This effect is conveniently studied by using 
Schwinger boson representation 
of the Heisenberg model \cite{assa}. On a bipartite lattice one represents spin
operators as
$ S_i^+ = a_i^{\dagger} b_i $ and $S_i^z = 1/2~ ( a_i^{\dagger} a_i - b_i^{\dagger} b_i ) $
on sublattice $A$, and $ S_i^+ = - b_i^{\dagger} a_i $ and 
$S_i^z = 1/2~ ( b_i^{\dagger} b_i - a_i^{\dagger} a_i ) $
on sublattice $B$. Provided that constraint $ a_i^{\dagger} a_i + b_i^{\dagger} b_i =1 $
is satisfied at each site and $a$ and $b$ obey the usual bosonic commutation relation we 
easily recover the proper commutation relations for spin $SU(2)$ algebra at each site.
In terms of $a$ and $b$ operators Hamiltonian (\ref{heisenberg})
can be written as 
\begin{eqnarray}
{\cal H} = - \frac{J}{2} \sum_{\langle ij \rangle} A_{ij}^{\dagger} A_{ij}
 + \frac{H_z}{2} \sum_i (-)^i ( a_i^{\dagger} a_i - b_i^{\dagger} b_i )
 + \sum_i \lambda_i (  a_i^{\dagger} a_i + b_i^{\dagger} b_i - 1 )
\end{eqnarray}
where $A_{ij} = a_i a_j + b_i b_j $ and the last terms enforces the constraint
at each site. In the mean field approximation the last Hamiltonian may be diagonalized
by quasiparticles $ \alpha_{a\mu, k} $ ($a=1,2$, $\mu=\pm$, and momentum $k$ 
runs in the magnetic zone only) 
with dispersion
$
\omega_{a\pm, k } = \omega_{k} \pm H_z/2
$, where
$
\omega_{k} = \sqrt{\lambda^2 - 4 Q^2 \gamma_k^2} 
$ and
$ \gamma_k =cos k_x + cos k_y $. Mean-field parameters
$\lambda$ and $Q$ have to be found from minimizing the free energy
\begin{eqnarray}
F = \frac{2}{\beta} \sum_k  ln \left\{ sinh[ \frac{\beta}{2} ( \omega_k 
	+ \frac{B_z}{2} )] \right\}
    + \frac{2}{\beta} \sum_k ln \left\{ sinh[ \frac{\beta}{2} ( \omega_k 
	- \frac{B_z}{2} )] \right\}
 - 2 \lambda N + \frac{ 4 Q^2}{J} N 
\label{F}
\end{eqnarray}
In terms of $\alpha$ operators the uniform and staggered spin operators can be written as
\begin{eqnarray}
M_{+} &=&  \frac{1}{N} \sum_k \left( \alpha_{1+,k}^{\dagger} \alpha_{2-,k} 
	- \alpha_{2+,k}^{\dagger} \alpha_{1-,k} \right) 
\hspace{0.65cm}
M_z   = \frac{1}{N} \sum_{a k} \left( \alpha^{\dagger}_{a+, k} \alpha_{a+, k} 
- \alpha^{\dagger}_{a-, k} \alpha_{a-, k} \right)
\label{M+}\\
N_z  &=& \frac{1}{N} \sum_{a \mu k} (-)^a \left\{ cosh( 2 \theta_k )  
	\alpha^{\dagger}_{a \mu, k} \alpha_{a \mu, k} 
+ \frac{1}{2} sinh ( 2 \theta_k ) [~ \alpha^{\dagger}_{a \mu, k} \alpha^{\dagger}_{a \mu, -k}
	+ \alpha_{a \mu, k} \alpha_{a \mu, -k}~] \right\} \\
N_x &=& \frac{1}{N} \sum_{\mu k}  \{~ cosh( 2 \theta_k )~ 
\alpha_{1\mu,k}^{\dagger} \alpha_{2\bar{\mu},k}
+ \frac{1}{2} sinh ( 2 \theta_k )~ [~\alpha_{1\mu,k}^{\dagger} \alpha_{2\bar{\mu},-k}^{\dagger} +
\alpha_{1\mu,k} \alpha_{2\bar{\mu},-k}~]~\}
\label{L}
\end{eqnarray}
where $tanh( 2 \theta_k ) = - 2 Q \gamma_k/ \lambda $.
From equation (\ref{M+}) we notice that the $M_+$ operator has the
exact energy of $H_z$ due to the fact that the mean-field approximation
on Schwinger bosons does not break spin symmetry. 
The correlation function for $N_z$ is given by
\begin{eqnarray}
D(\omega>0)\! =\! Im\! \int\! dt e^{i \omega t} \theta(t) \langle N_z(t) N_z(0) \rangle
= \! \frac{1}{N}\! 
\sum_k \! \left( 1 + N(\omega_{+,k}) + N(\omega_{-,k}) \right) \delta ( \omega \!- \!\omega_{+,k} 
\!- \!\omega_{-,k} )
\end{eqnarray}
where $N(\omega) = 1/(e^{\beta \omega}-1)$. 
At $T<T_c$, bosons
$\alpha_{1-,k=0}$ and $\alpha_{2-,k=0}$  are condensed
so there is a $\delta$-function resonance in $ D(\omega) $ at frequency $ H_z = 2 \mu$  
with the weight $sinh^2( 2 \theta_{k=0} ) = \langle N_x^2 \rangle
/ \langle M_z \rangle$ as dictated by the $SU(2)$ symmetry \cite{zhang-yang,eta,ijmp}.
There is also an additional broad peak due to $\alpha_{a-,k \neq 0}$ that starts at energy
$ 2 \mu$ and then extends in a frequency range of around $T$ with an integrated weight proportional
to $T^{3/2}$. Above the Neel ordering temperature the $\delta$-peak in $D(\omega)$ is absent, 
however there is  a broadened peak at the energy 
slightly larger than $2 \mu$ (the lower energy threshold 
is given by $ \omega_{min} = 
\omega_{-,k=0} + \omega_{+,k=0} = 2 \mu + 2 \omega_{-,k=0} $, and $ \omega_{-,k=0} $
is small since it vanishes at $T_c$ ). The width of this broad
feature is $T$ and the total spectral weight is proportional to 
$T^{3/2} e^{ -  \omega_{-,k=0} /T}$.  As the temperature is lowered
and the system approaches Bose condensation at $T_c$, the energy $ \omega_{-,k=0}$ becomes
smaller leading to a smooth increase of the integrated intensity of the peak.
So the intensity of the resonance changes continuously across $T_c$ 
but as may be shown by a detailed analysis there is a jump in the derivative. 
On Figure \ref{fig2} we show  a sketch of $ D(\omega) $ 
for temperatures below and above $T_c$.  

Before concluding we would like to remark that
that many features of the precursors of the $\eta$ excitations that we discussed here,
such as temperature dependent shift to higher energies and considerable broadening
are likely to be present for the $\pi$-excitation of the $t$-$J$ model as well.
INS on underdoped $YBa_2Cu_3O_{6+\delta}$ reveals considerable
broadening of the resonance above $T_c$, however the current accuracy 
of experiments does not allow to see any changes in the energy of the 
resonance above the superconducting transition.

This work is partially supported by the NSF at ITP (ED) and by the NSF grant 
DMR-9814289. We acknowledge useful discussions with 
A. Auerbach and D.J. Scalapino.


\begin{thebibliography}{10}

\bibitem{optimal} 
H.~Mook {\it et. al.}, {\em Phys. Rev. Lett.} 
{\bf 70,} 3490 (1993); 
J. Rossat-Mignod {\it et. al.},  {\em Physica C}, 
{\bf 235} 59 (1994); 
H. Fong {\it et. al.}, {\em Phys. Rev. Lett.} 
{\bf 75,} 316 (1995); 

\bibitem{underdoped} 
P.~Dai {\it et. al.}, {\em Phys. Rev. Lett.} 
{\bf 77,} 5425 (1997);
H. Fong {\it et. al.}, {\em Phys. Rev. Lett.} 
{\bf 78,} 713 (1997); 

\bibitem{BSCO}
H. Mook {\it et. al.}, cond-mat/9811100;
H. Fong {\it et. al.}, cond-mat/9902262.


\bibitem{theories}
N.~Bulut and D.~Scalapino, {\em Phys. Rev. B} 53, 5149 
  (1996);
I.~Mazin and V.~Yakovenko, {\em Phys. Rev. Lett.} 75, 4134 
(1995);
D.Z.~Liu {\it et al.}, {\em Phys. Rev. Lett.}
  75, 4130 (1995);
F.~Onufrieva {\it et al.}, {\em Physica} (
  Amsterdam ) 251C,  348 (1995);
G.~Blumberg {\it et al.}, {\em
    Phys. Rev. B} 52, 15741 (1995);
Y.~Zha {\it et al.},  {\em Phys. Rev. B }
  54, 7561 (1996); 
A.J.~Millis and H.~Monien, cond-mat/9606008;
L.~Yin {\it et al.},
  cond-mat/9606139; P.W.~Anderson, J. Phys. Condens. Matter 8, 10083
  (1996); 
A.~Abrikosov, preprint.


\bibitem{prl95} E.~Demler and S.C.~Zhang,
{\em Phys. Rev. Lett.}, 75:4126, 1995.

\bibitem{meixner} S. Meixner {\it et. al.}, {\em Phys. Rev. Lett.} 
{\bf 79,} 4902 (1997). 

\bibitem{prb} 
E. Demler, H. Kohno and S.C. Zhang, {\em Phys. Rev. B} 
{\bf 58,} 5719 (1998). 


\bibitem{eder} R.~Eder, W. Hanke and S.C. Zhang,  
{\em Phys. Rev B}. {\bf 57}, 13781 (1998). 


\bibitem{scalapino}
D.J. Scalapino and S. White , {\em Phys. Rev. B} 
{\bf 58,} 8222 (1998). 


\bibitem{nature}
E.~Demler and S.C.~Zhang, {\em Nature } 396, 733 (1998).

\bibitem{helium}
P.W. Anderson and W.F. Brinkman, {\em Phy. Rev. Lett.} {\bf 30}, 1108 (1973).



\bibitem{loram}
J. Loram {\it et al},  {\em Physica}  {\bf C 171},  243-256 (1990);  
J. Loram {\it et al},  {\em Journal of Superconductivity} {\bf 7},  
243 (1994).

\bibitem{science}
S.C. Zhang, {\em Science}, {\bf 275} 1089-1096 (1997).


\bibitem{scz_proc} Shou-Cheng Zhang, cond-mat/9808309.


\bibitem{phase_fluct} 
S. Doniach and M. Inui, {\em Phys. Rev. B} {\bf 41}, 6668 (1990);
V. Emery and S. Kivelson, {\em Nature} {\bf 374}, 434 1995;
M. Randeria, {\em  J. Phys. Chem. Solid} {\bf 59}, 10 (1998).



\bibitem{yang} 
C.N. Yang {\em Phys. Rev. Lett.} 
{\bf 63,} 2144 (1989). 

\bibitem{zhang-yang}
C.N. Yang and S.C. Zhang, {\em Mod. Phys. Lett. } {\bf B 4}, 759 (1990).

\bibitem{eta} S. C. Zhang,
{\em Phys. Rev. Lett.} {\bf 65}, 120 (1990).


\bibitem{assa}
A. Auerbach, Interacting electrons and quantum magnetism, {\em Springer-Verlag, New York},
 (1994)

\bibitem{ijmp}
E. Demler, S.C.~Zhang, N.~Bulut, and D.J.~Scalapino, 
 {\em Int. Journal of Modern Physics} B, {\bf 10}, 2137 (1996)




\end{thebibliography}
 \end{document}